\documentclass[HYPER]{JHEP}
\usepackage{graphicx,amssymb,bm,latexsym,amsmath}
\usepackage{epstopdf}
\usepackage{caption}
\usepackage{subcaption}
\usepackage[section]{placeins}
\usepackage{float}
\usepackage{capt-of}
\usepackage{tabu}
\usepackage[toc,page]{appendix}
\newcommand{\el}{\nonumber}
\newcommand{\mathsym}[1]{{}}
\newcommand{\unicode}[1]{{}}

\title{Chaos in classical string dynamics in $\hat{\gamma}$ deformed $AdS_5 \times T^{1,1}$}
\author{Kamal L. Panigrahi \\
Department of Physics, Indian Institute of Technology Kharagpur, \\ Kharagpur-721 302, India \\
{\rm and} Theory Group-DESY, Hamburg, Notkestrasse 85, \\ D-22603 Hamburg, Germany\\
Email:\email{panigrahi@phy.iitkgp.ernet.in}}
\author{Manoranjan Samal\\
Department of Physics, Indian Institute of Technology Kharagpur, \\
Kharagpur-721 302, India\\
Email: \email{manoranjan@phy.iitkgp.ernet.in}}
\abstract{We consider a circular string  in $\hat{\gamma}$ deformed $AdS_5 \times T^{1,1}$ which is localized in the center of $AdS_5$ and 
winds around the two circles of  deformed $T^{1,1}$. We observe chaos in the phase space of the circular string implying non-integrability 
of string dynamics. The chaotic behaviour in phase space is controlled by  energy as well as the deforming parameter $\hat{\gamma}$.
We further show that the point like object exhibits  non-chaotic behaviour. Finally we calculate the Lyapunov exponent for both extended and 
point like object in support of our first result.}
\begin{document}
\section{Introduction}
The AdS/CFT correspondence is a powerful technique that provides an interplay between the gauge theory without gravity and a string theory
(supergravity theory) with gravity\cite{M},\cite{W},\cite{GKP}. The most studied example is the duality between type IIB string theory on $AdS_5 \times S^5$ and 
${\cal N}$=4 supersymmetric Yang Mills (SYM) theory in $D=4$. It is particularly well understood in the strong coupling limit of the field theory side. 
In this connection integrability on both sides of the duality has played a key role in the understanding the duality better. In particular it has 
helped us in getting close to a solution of ${\cal N}=4$ SYM in the planar limit \cite{review}. The fact that both sides of the duality are integrable in the planar limit leaves us 
interested in looking at the theories more closely.  Over the past few years there has been enormous amount of work devoted towards the advancement of integrability and that in turn has opened up the possibility of looking the integrability techniques in a much wider context, e.g. looking 
the theories, beyond the planar limit and in the background of deformationed of AdS. In this context the semiclassical strings have played very important role.  
Semiclassical quantization is one of the most popular approach to probe general string backgrounds with various background fluxes. 
Classical solutions and
trajectories of rotating strings, and D-branes, have played an important role in understanding the AdS/CFT correspondence which was otherwise
obscure at times.  Semiclassical quantization has played a vital role in  the study of BMN, \cite{BMN}, GKP \cite{GKP} and rigidly rotating strings
\cite{Frolov:2002av} which can all be 
understood as classical trajectories of the the rotating and pulsating string. These classical trajectories have been one of the main ingredients of the
understanding of the semiclassical AdS/CFT from the string theory prospective. In general, the string dynamics in curved space are described 
by the help of 2d sigma models where equations of motion in general are non-linear. Integrability plays an major role in finding out the  
classical solutions of the  non linear equations, correlation functions, scattering amplitudes and spectrum. Therefore it is important 
to check the integrability of string sigma model in a specific background. 
 
In the context of integrability on the other hand, it is a common  fact that the phase space of most mechanical systems is not integrable and 
thus the role of chaotic classical trajectories has been investigated in  detail in the past. In general a system is said to be integrable if the 
number of degrees of freedom is same as the number of conserved charges. String sigma model in two dimensions has infinite number of degrees of freedom and 
the system is  integrable on arbitrary backgrounds only
when it has infinite number of conserved charges which happens to be the case in $AdS_5 \times S^5$ \cite{Bena:2003wd}. The standard way to show the integrability 
of 2d-sigma model in arbitrary background is to construct a lax pair which generates infinite number of conserved charges. 
But to show the existence of Lax pair is quite comber some. Infact the necessary condition for a system to be integrable is when all of its subsystem
are integrable. In other words, a system is said to be non-integrable if at least one of its subsystem is non-integrable. Therefore the general proof 
of the non integrability of a two dimensional sigma model in arbitrary background can be done by first reducing it to a 1d subsystem 
and then showing the 1d subsystem is non-integrable. This can be done either by doing numerical analysis of string motion in phase 
space or by analytic method using normal variational equation(NVE).  This numerical approach has been particularly useful in various 
cosmological and black hole backgrounds. 
Using numerical method it has been shown that phase space of a test circular cosmic string in Schwarzschild black hole geometry is chaotic
\cite{Larsen:1993nt},\cite{Frolov:1999pj},\cite{Zayas:2010fs}. It has further been found analytically  that the Friedmann-Robertson-Walker (FRW)
cosmological model is completely integrable only for some special value of the cosmological constant \cite{Boucher:2007zz}. 
The evidence of chaotic behaviour has been noticed in $AdS_5 \times T^{1,1} $ background \cite{T11} and in its 
Penrose limit \cite{Asano:2015qwa}. Applying the analytic technique it has been shown the $AdS_5 \times X^5$ geometries are non-integrable,
where $X^5$ is in a general class of five-dimensional Einstein spaces \cite{Basu:2011fw}. In case of non-relativistic theories it has been shown the
integrability nature depends on dynamical critical exponent \cite{Giataganas:2014hma},\cite{Bai:2014wpa}. 
Taking classical spinning string solution in various supergravity backgrounds \cite{Basu:2012ae} it has been shown the phase spaces 
are chaotic and hence non-integrable. More recently the integrability of curved brane backgrounds has been studied and it is 
found except for some specific limit the extended string motion is non-integrable while the point like string dynamics is always integrable
\cite{D-brane}.  Apart from these there are number of instances where the integrability is studied by the help of either analytical method or numerical analysis\cite{Basu:2011dg},\cite{complex-beta},\cite{Asano:2015eha}. Motivated by the recent interest in studying the classical integrability of
string motion in various backgrounds and its connection with chaotic motion of the test string in generic deformed background and otherwise,
we study the motion of classical circular string in $\hat{\gamma}$ deformed $AdS_5 \times T^{1,1}$ background. We have shown numerically 
the appearance of chaos for a circular string moving in the deformed background. 
 
The rest of the paper is organised as follows. In section 2 we write down $\hat{\gamma}$ deformed $AdS_5 \times T^{1,1}$ background 
geometry and the fields. In section 3 we study
a consistent string sigma model, taking a semi-classical circular string ansatz. We write down the equations of motion for the test string
for the given ansatz and construct all the conserved charges. Section 4 is devoted to the study of chaos in the classical string dynamics 
by two different techniques, namely first by looking at the Poincare section and then by studying the Lyapunov exponent. 
Finally, in section 5 we conclude with some comments. 
 
\section{The ${\hat\gamma}$ deformed $ \mathbf{AdS_5 \times T^{1,1}}$ background}
The $AdS_5 \times T^{1,1}$ geometry is the gravity dual of $\mathcal{N}=1$ super symmetric Yang-Mills theory, which arises 
from the near horizon geometry of a stack of N number of D3-branes at the tip of the conifold, where the base of the conic is $T^{1,1}$. 
The metric of $AdS_5 \times T^{1,1}$ is 
given by
\begin{align}
ds^2 &=ds_{AdS_5}^2+ds_{T^{1,1}}^2 \el \\
ds_{AdS_5}^2 &=-\cosh^2 \rho dt^2 +d\rho^2+\sinh^2 \rho d\Omega_3^2 \el \\
ds_{T^{1,1}}^2 &=\frac{1}{6}\sum_{i=1}^2\left[d\theta_i^2+\sin^2\theta_i d\phi_i^2\right]+\frac{1}{9}\left[ d\psi +\cos \theta_1 d\phi_1 +\cos\theta _2 d \phi_2\right]^2.
\end{align}
The internal manifold $T^{1,1}$ is a five dimensional Sasaki-Einstein manifold and is the coset space $(SU(2)\times SU(2))/U(1)$.
Applying the TsT transformation to this gives rise to the so called $\hat{\gamma}$ deformed $AdS_5 \times T^{1,1} $ metric and NS-NS two 
forms $(b_{mn})$ \cite{Lunin:2005jy},\cite{CatalOzer:2005mr}.
\begin{align}
ds^2 &= ds^2_{AdS}+G(\hat{\gamma}) \left[\frac{1}{6}\sum_{i=1}^2(G^{-1}(\hat{\gamma})d\theta_i^2+\sin^2 \theta_i d\phi_i^2) \right. \el \\
&+\left. \frac{1}{9}(d\psi+\cos \theta_1 d\phi_1+\cos \theta_2  d\phi_2)^2+\hat{\gamma}^2\frac{\sin^2 \theta1 \sin^2 \theta_2}{324}d\psi^2 \right].
\end{align}
\begin{align}
b_{mn}&= \hat{\gamma}G(\hat{\gamma})\left[ \frac{\cos \theta_2 \sin ^2 \theta_1}{54} d \phi_1 \wedge d\psi-\frac{\cos \theta_1 \sin ^2 \theta_2}
{54} d\phi_2 \wedge d\psi \right. \el \\
&+ \left. \left(\frac{\sin^2 \theta_1 \sin ^2 \theta_2 }{36}+ \frac{\cos^2 \theta_1 \sin^2 \theta_2+ \cos ^ 2\theta_2 \sin^2 \theta_1 }{54}\right)d\phi_1 
\wedge d \phi_2\right],
\end{align}
where
\begin{align*}
G(\hat{\gamma} )^{-1}&\equiv 1+ \hat{\gamma}^2 \left(\frac{\sin^2 \theta_1 \sin ^2 \theta_2 }{36}+\frac{\cos^2 \theta_1 \sin^2 \theta_2
+ \cos ^ 2\theta_2 \sin^2 \theta_1 }{54}\right).
\end{align*}
The above deformed geometry has also been achieved by making a deformation of classical Yang-Baxter sigma model as described 
in \cite{Crichigno:2014ipa}.
\section{The string sigma-model and circular string} In this section we shall start our analysis by making the following ansatz for the circular string
\begin{align}
\rho= 0,   ~~~~~ \theta_i=\theta_i(\tau), ~~~~~ \phi_1=\alpha_1 \sigma,  ~~~~ \phi_2=\alpha_2 \sigma, ~~~~~ \psi= \psi(\tau).
\end{align}
It shows that the string is localized at the center of the AdS whereas it extends along
the two angles($\phi_1$, $\phi_2$) of deformed $T^{1,1}$  with winding numbers $\alpha_1$ and $\alpha_2$ respectively. Here we have chosen 
such type of ansatz because  we can truncate 2d sigma-model to 1d dynamical Hamiltonian system and the same time we can study its
dynamics in phase space. The 2d sigma-model action in generic background is written as  
\begin{align}
S= -\frac{1}{4 \pi \alpha\prime}\int d\tau d\sigma \left [ \sqrt{-h}h^{\alpha \beta}g_{mn}\partial _\alpha x^m \partial_ \beta x^n 
-  {\epsilon}^ {\alpha \beta }\partial _\alpha x^m \partial_ \beta x^n b_{mn} \right] ,
\end{align}
where $m, n$ are the spacetime indices. Further in conformal gauge $h^{\alpha \beta}$=diag(-1,1) and as usual $\epsilon^{\tau \sigma}=-\epsilon^{\sigma \tau}=1$.
Now we can write the Lagrangian from the action as
 \begin{align}
\mathcal{L} &= -\frac{\dot{t}^2}{2}+\frac{1}{12}(\dot{\theta}_1^2 + \dot{\theta}_2^2)-\frac{G(\hat{\gamma})}{36}(\alpha_1^2 \sin^2\theta_1+\alpha_2^2 \sin^2 \theta_2)-\frac{G(\hat{\gamma})}{18}(\alpha_1^2+\alpha_2^2) \el \\
&\hspace{5mm}-\frac{G(\hat{\gamma})}{9}\alpha_1 \alpha_2 \cos\theta_1 \cos \theta_2 +\dot{\psi}^2 G(\hat{\gamma})\left( \frac{1}{18}+\frac{\hat{\gamma}^2 \sin ^2 \theta_1 \sin^2 \theta_2}{648}\right) \el \\
&+\frac{\hat{\gamma}G(\hat{\gamma})}{54}\left( \alpha_2\cos \theta_1 \sin^2 \theta_2-\alpha_1 \cos \theta_2 \sin^2 \theta_1\right)\dot{\psi}.
\end{align}
The canonical momenta are introduced as
\begin{align}
p^{\tau}_m=\frac{\partial \mathcal{L}}{\partial(\partial_\tau x^m)}= \sqrt{-h} h^{\tau \alpha } \partial_\alpha x^ng_{mn}-\epsilon^{\tau  \beta} \partial _\beta x^n b_{mn}.
\end{align}
Using canonical momenta and Lagrangian density we can get the Hamiltonian density,
\begin{align}
\mathcal{H}&=-\frac{E^2}{2}+3(p_{\theta_1}^2+p_{\theta_2}^2)+\frac{G(\hat{\gamma})}{36}(\alpha_1^2 \sin^2\theta_1+\alpha_2^2 \sin^2 \theta_2)+\frac{G(\hat{\gamma})}{9}\alpha_1 \alpha_2 \cos\theta_1 \cos \theta_2 \el \\
& \hspace{15mm}+\frac{G(\hat{\gamma})}{18}(\alpha_1^2+\alpha_2^2)+\frac{\left( J -\frac{\hat{\gamma}G(\gamma)}{54}(\alpha_2\cos \theta_1 \sin^2 \theta_2-\alpha_1 \cos \theta_2 \sin^2 \theta_1)\right)^2}{2 G(\hat{\gamma})\left( \frac{1}{9}+\frac{\hat{\gamma}^2 \sin ^2 \theta_1 \sin^2 \theta_2 }{324}\right)}. 
\end{align}\label{eoms}
Variation of action with respect to $x^m$ gives the following equation of motion,
\begin{align}
2\partial_\alpha(\sqrt{-h}h^{\alpha \beta}g_{km}\partial_\beta x^m)-&\sqrt{-h}h^{\alpha \beta} \partial_kg_{mn}\partial_\alpha x^m \partial_\beta x^n-2\partial_\alpha \epsilon^{\alpha \beta}\partial_\beta x^m b_{km}\el \\+&\epsilon^{\alpha \beta}\partial_\alpha x^m \partial_\beta x^n \partial_k b_{mn}=0 
\end{align}
Further,the variation of action with respect to metric  gives the Virasoro constraints,
\begin{align} \label{virasoro}
g_{mn}(\partial_\tau x^m \partial _\tau x^n+\partial_\sigma x^m \partial _\sigma x^n)=0  \\ 
g_{mn}(\partial_\tau x^m \partial _\sigma x^n)=0.
\end{align}
The equations of motion for $t$ and $\psi$ leads, respectively, to
\begin{align}
&\dot{t}=E, \\
&\dot{\psi} G(\hat{\gamma})\left(\frac{1}{9}+\frac{\hat{\gamma}^2 \sin^2 \theta_1 \sin^2 \theta_2}{324}\right)+\frac{\hat{\gamma}G(\hat{\gamma})}{54}\left( \alpha_2\cos \theta_1 \sin^2 \theta_2-\alpha_1 \cos \theta_2 \sin^2 \theta_1\right)=J ,
\end{align}
here $E$ and $J$ both are constants motion. Further, the equations motion of $\theta_1$ and $\theta_2$ are non-trivial and are given by,
\begin{align}
&\ddot{\theta_1} =-G(\hat{\gamma})\left(\frac{1}{3}\alpha_1 ^2 \cos\theta_1 \sin \theta_1- \frac{2}{3} \alpha_1 \alpha_2 \cos\theta_2 \sin \theta_1- \frac{\hat{\gamma}^2 \sin 2 \theta_1 \sin ^2 \theta_2 \dot{\psi}^2}{108}\right. \el \\
&\left.+ \frac{\hat{\gamma} }{9}(\alpha_2 \sin ^2 \theta_2 \sin \theta_1+\alpha_1 \cos \theta_2 \sin 2 \theta_1)\dot{\psi}\right)- G_{\theta_1} F
\end{align}
\begin{align}
\ddot{\theta_2} &=-G(\hat{\gamma})\left(\frac{1}{3}\alpha_2^2 \cos\theta_2  \sin\theta_2-\frac{2}{3} \alpha_1 \alpha_2 \cos\theta_1 \sin\theta_2 -\frac{\hat{\gamma}^2 \sin 2 \theta_2 \sin ^2 \theta_1 \dot{\psi}^2}{108}\right. \el \\ &\left.-\frac{\hat{\gamma} }{9}(\alpha_1 \sin ^2 \theta_1 \sin \theta_2 +\alpha_2 \cos \theta_1 \sin 2 \theta_2)\dot{\psi}\right)- G_{\theta_2} F
\end{align}
where

\begin{align*}
G_{\theta_1}=-\frac{54 \hat{\gamma}^2 (3+\cos 2\theta_2)\sin 2 \theta_1}{(108+2 \hat{\gamma }^2\cos^2 \theta_2 \sin^2 \theta_1 +\hat{\gamma }^2\cos^2 \theta_1 \sin^2 \theta_2+3 \hat{\gamma }^2\sin^2 \theta_1\sin^2 \theta_2)},
\end{align*}

\begin{align*}
G_{\theta_2}=-\frac{54 \hat{\gamma}^2 (3+\cos 2\theta_1)\sin 2 \theta_2}{(108+2 \hat{\gamma }^2\cos^2 \theta_1 \sin^2 \theta_2 +\hat{\gamma }^2\cos^2 \theta_2 \sin^2 \theta_1+3 \hat{\gamma }^2\sin^2 \theta_2\sin^2 \theta_1)},
\end{align*}
and
\begin{align*}
F&=\frac{1}{3}(\alpha_1^2 +\alpha_2^2)+\frac{2}{3} \alpha_1 \alpha_2 \cos \theta_1 \cos \theta_2+\frac{1}{6}(\alpha_1^2\sin^2 \theta_1+\alpha_2^2 \sin^2 \theta_2) -\dot{\psi}^2\left( \frac{1}{3}+\frac{\hat{\gamma}^2 \sin ^2 \theta_1 \sin^2 \theta_2}{108}\right) \el \\ &~~~~~~~-\frac{\hat{\gamma}}{9}\left( \alpha_2\cos \theta_1 \sin^2 \theta_2-\alpha_1 \cos \theta_2 \sin^2 \theta_1\right)\dot{\psi}.
\end{align*}
From \ref{virasoro}  the Virasoro constraints can be written as,
\begin{align} \label{conseq}
E^2&=\frac{1}{6}(\dot{\theta_1}^2+\dot{\theta_2}^2)+\frac{G(\hat{\gamma})}{18}(\alpha_1^2 \sin^2\theta_1+\alpha_2^2 \sin^2 \theta_2^2)+2\frac{G(\hat{\gamma})}{9}\alpha_1 \alpha_2 \cos\theta_1 \cos \theta_2 \el \\
& \hspace{15mm}+\frac{G(\hat{\gamma})}{9}(\alpha_1^2+\alpha_2^2)+\frac{\left( J -\frac{\hat{\gamma}G(\gamma)}{54}(\alpha_2\cos \theta_1 \sin^2 \theta_2-\alpha_1 \cos \theta_2 \sin^2 \theta_1)\right)^2}{ G(\hat{\gamma})\left( \frac{1}{9}+\frac{\hat{\gamma}^2 \sin ^2 \theta_1 \sin^2 \theta_2 }{324}\right)}.
\end{align} 
The Hamiltonian is fixed to zero by Virasoro constraints. Since the equations of motion $\theta_1$ and $\theta_2$ are complicated, it is very difficult to find out normal variational  equation(NVE) and study the integrability analytically. We will study the problem from
a numerical analysis by showing the appearance of chaos in the next section.
\section{Numerical analysis}
The non integrability nature of the string dynamics can be verified numerically by showing chaotic behavior of the string  in its phase space. There are various techniques to show the chaotic behavior of the string dynamics in a particular background. Here we have used two methods. First we wish to study it from the point of view of Poincar\'{e} section  and then calculating the Lyapunov Exponent.
\subsection{Poincar\'{e} Section}
The string trajectories in phase space are distorted torus or the famous Kolmogorov-Arnold-Moser (KAM) torus [Fig.1]. 
As time evolves, the trajectories wind over the torus shows 
the quasi-periodic nature of the trajectories. It will be convenient to take the projection of the trajectories over a surface for studying their dynamics. These projections over a surface is called Poincar\'{e} section or surface of section \cite{Strogatz},\cite{ott}.
\begin{figure}[H]
\begin{center}
\includegraphics[width=.5\textwidth]{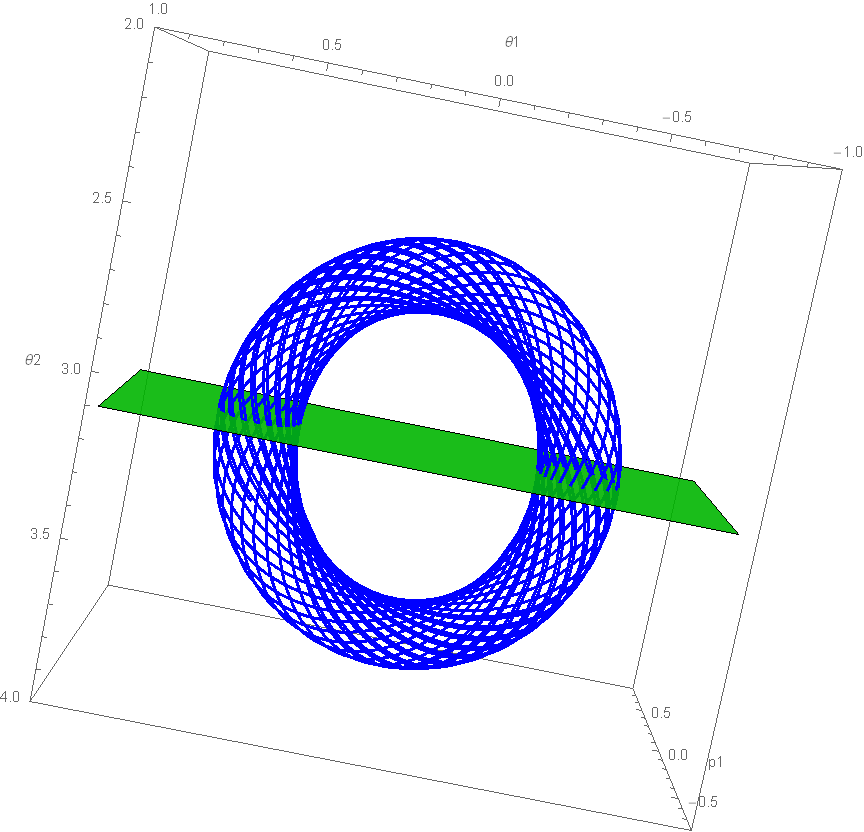}
    \caption{}
    \label{fig:f1}
\end{center}
\end{figure}
In the present case, the system has four phase space coordinates  ($\theta_1,\theta_2,p_{\theta_1},p_{\theta_2}$). 
The Virasoro constraint reduces it to a three dimensional subspace.  Different initial conditions to the phase space coordinates give 
different tori in the phase space. To find  different set of initial conditions we take $p_{\theta_1}=0,\theta_2=\pi $, then keeping energy a 
constant in  equation \ref{conseq} we vary $\theta_1 $ to get the corresponding values of $p_{\theta_2}$.  For an extended string we take 
the winding numbers of both $\theta_1$ and $\theta_2$ coordinates to 1. The intersection of the tori with the surface $\sin \theta_2=0$ 
gives distorted circles.\\
When energy is small these tori in the phase space are distinct [Fig.2(a)]. Each colour correspond to different set of initial conditions.  
As the energy of the system increased gradually some of the tori get deformed and destroyed by making a collection of scattered 
points in the phase space [fig.2(b)-2(d)]. These distorted tori are called cantori. At some higher value of energy  all the tori  get distorted 
and phase space become chaotic[Fig.2(e)].
\begin{figure}[H]
    \centering
    \begin{subfigure}[t]{0.5\textwidth}
        \centering
        \includegraphics[width=1\textwidth]{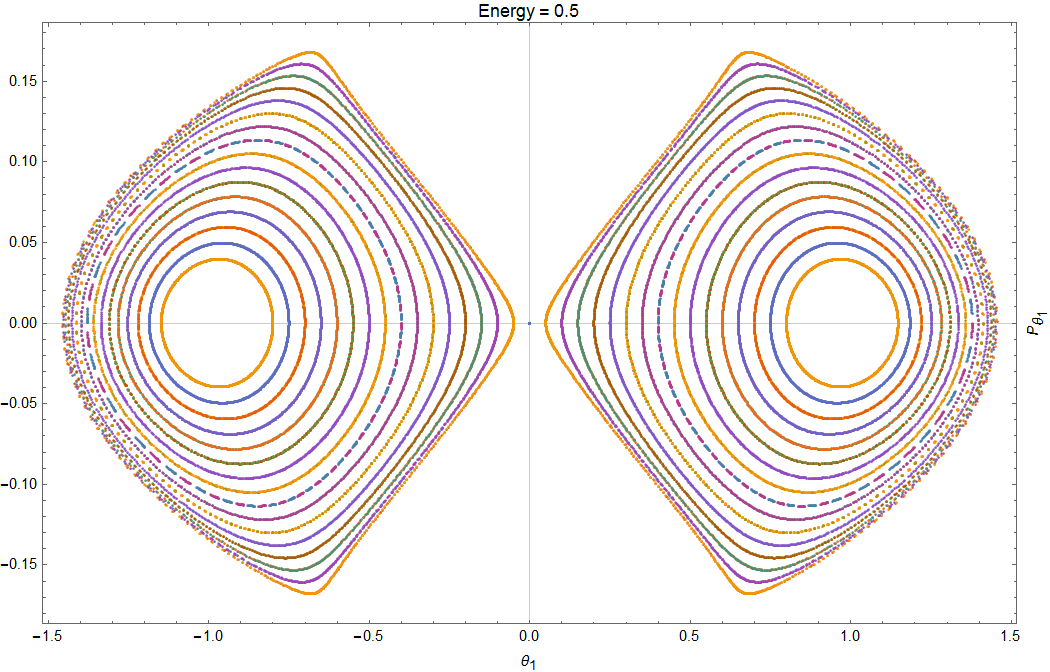}
        \caption{}
    \end{subfigure}%
    ~ 
    \begin{subfigure}[t]{0.5\textwidth}
        \centering
        \includegraphics[width=1\textwidth]{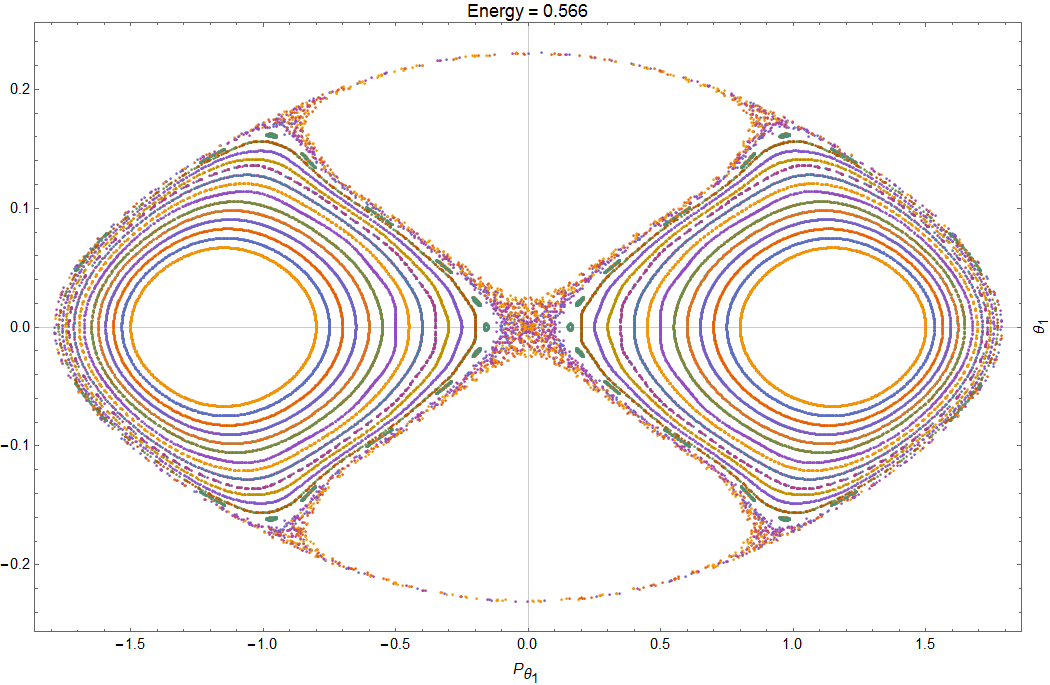}
        \caption{}
    \end{subfigure}
    \newline
    \begin{subfigure}[t]{0.5\textwidth}
        \centering
        \includegraphics[width=1\textwidth]{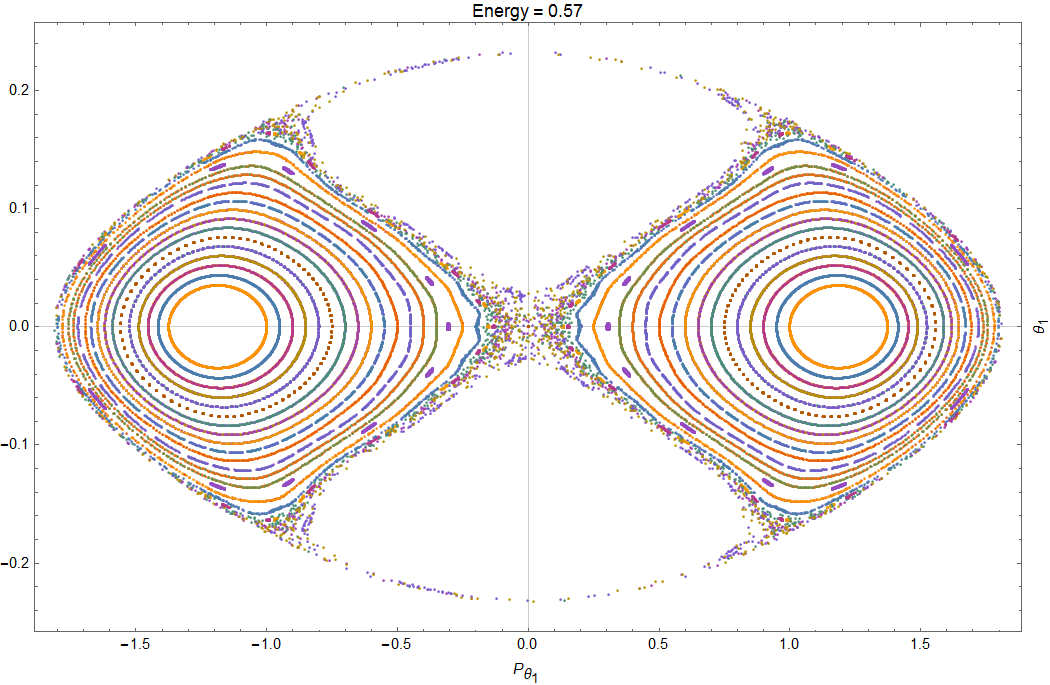}
        \caption{}
    \end{subfigure}%
    ~ 
    \begin{subfigure}[t]{0.5\textwidth}
        \centering
        \includegraphics[width=1\textwidth]{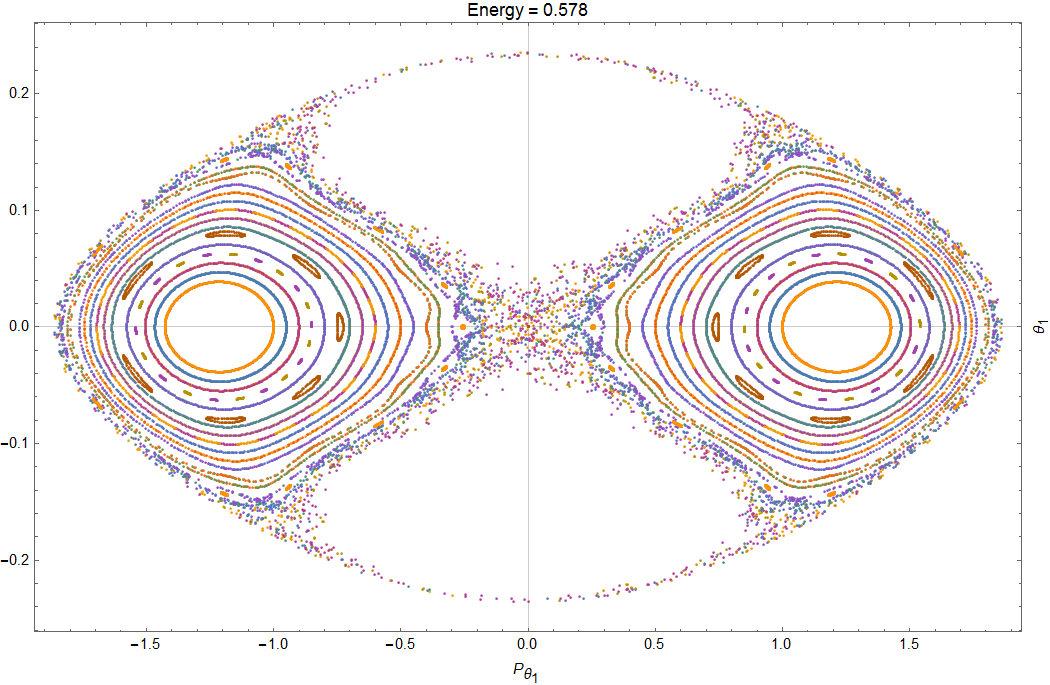}
        \caption{}
    \end{subfigure}
    \begin{subfigure}[t]{0.5\textwidth}
        \centering
        \includegraphics[width=1\textwidth]{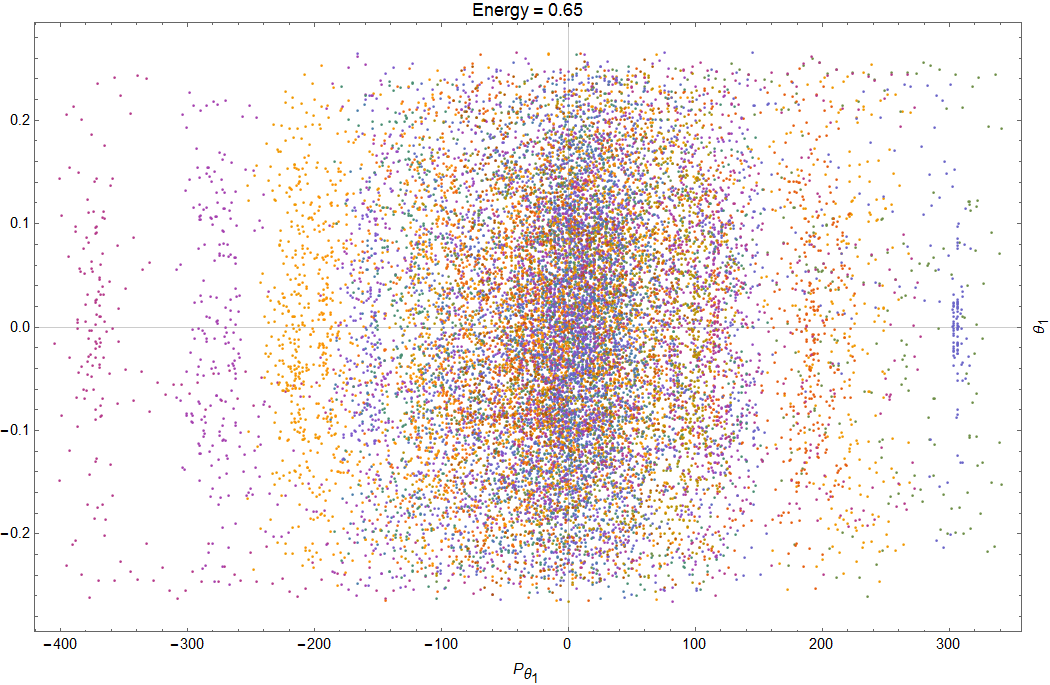}
        \caption{}
    \end{subfigure}
    \caption{Poincar\'{e} section for $\gamma=1$ }
\end{figure}
\clearpage
In Fig.3 it is observed that chaotic nature of phase space not only depends on energy but also the deformation parameter $(\hat\gamma)$ of 
$AdS_5 \times T^ {1,1} $ background. Keeping energy constant $E=0.5$ when $\hat{\gamma} $ is changed, the tori 
get distorted as earlier case and becomes chaotic for a higher value of $\hat{\gamma}$.

\begin{figure}[H]
    \centering
    \begin{subfigure}[t]{0.5\textwidth}
        \centering
        \includegraphics[width=1\textwidth]{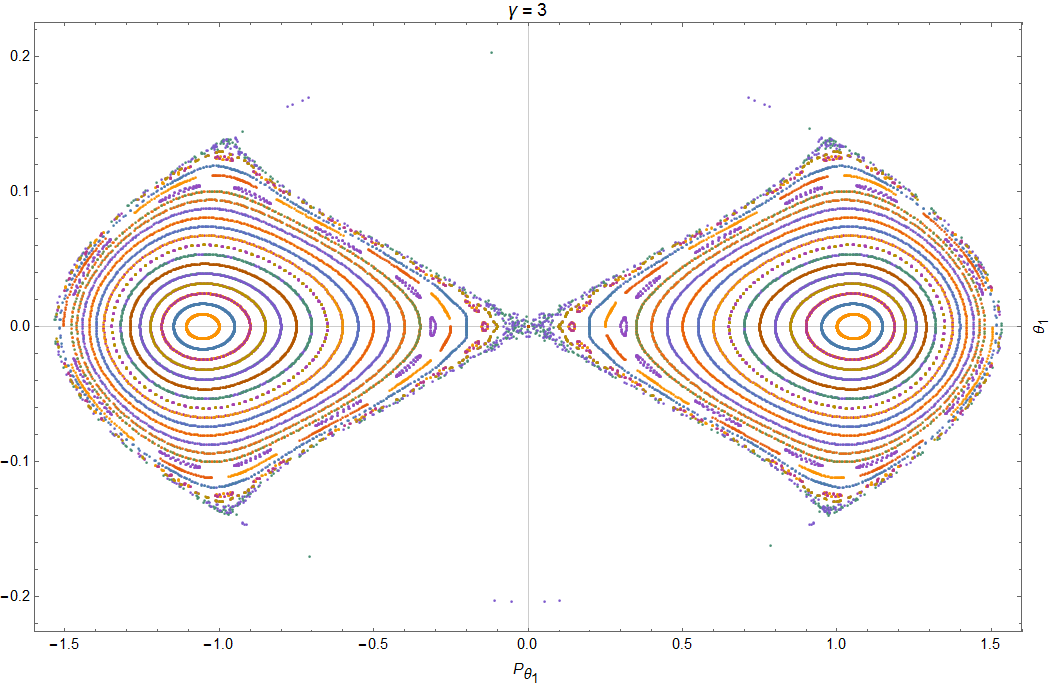}
        \caption{}
    \end{subfigure}%
    ~ 
    \begin{subfigure}[t]{0.5\textwidth}
        \centering
        \includegraphics[width=1\textwidth]{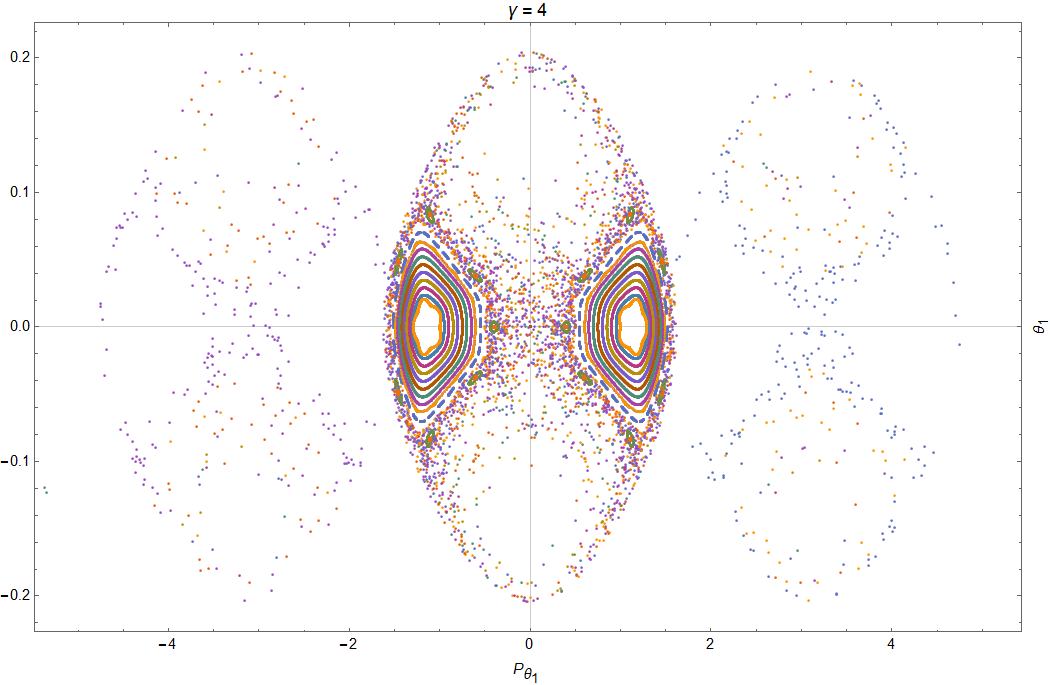}
        \caption{}
    \end{subfigure} 
    \newline
    \begin{subfigure}[t]{0.5\textwidth}
        \centering
        \includegraphics[width=1\textwidth]{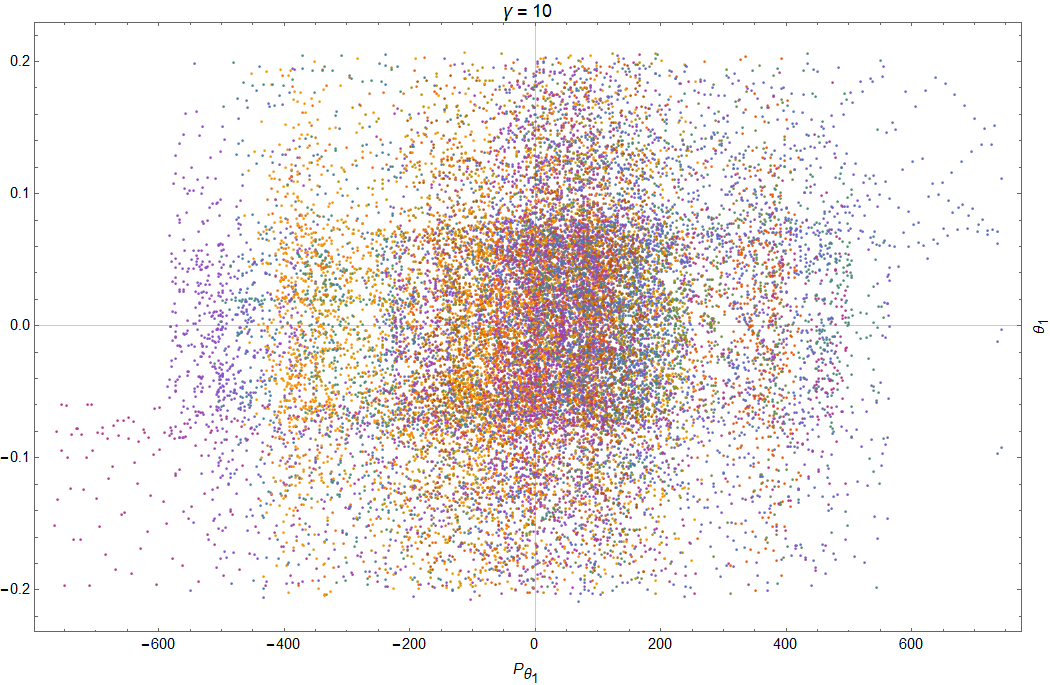}
        \caption{}
    \end{subfigure}%
    \caption{}
    \end{figure}
  \underline{\bf Point like string}\\
Let us look at the fate of the point like string in the deformed background. If we make the winding numbers tends to zero then the string is no longer
extended and it becomes point like. We  observe the phase space of a point like string is ordered and non-chaotic [fig.4]. This ordered behaviour of phase space remains unchanged even with the varying energy or the deformation parameter $\hat{\gamma}$.
 \begin{figure}[H]
\begin{center}
\includegraphics[width=.5\textwidth]{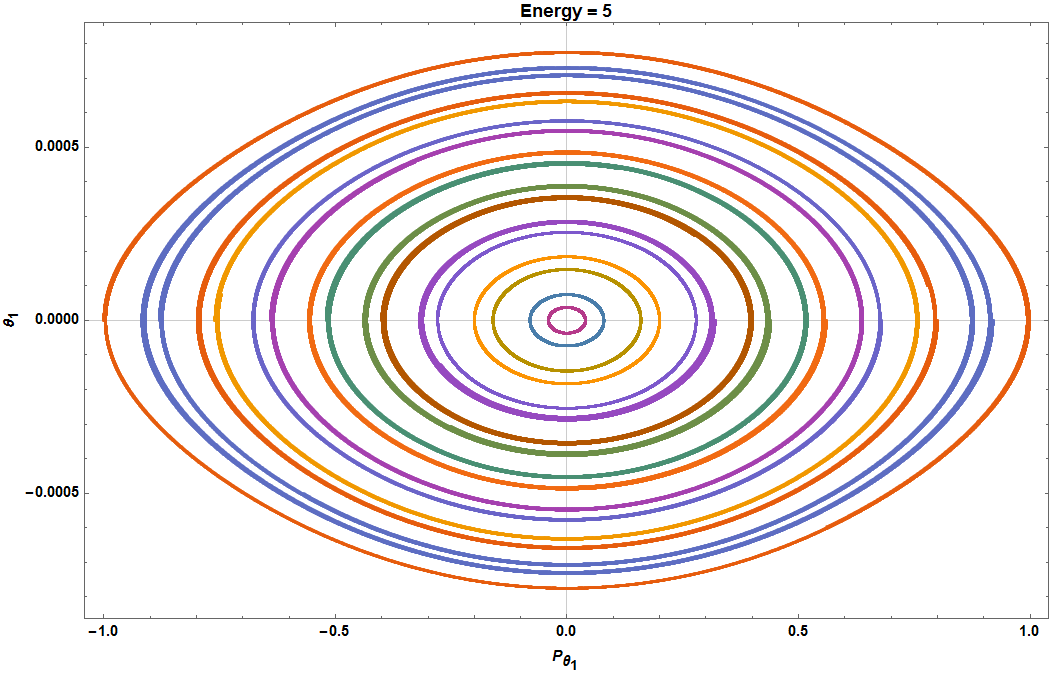}
    \caption{}
    \label{fig:f1}
\end{center}
\end{figure}
\subsection{Lyapunov Exponent}
Chaotic nature of the trajectories can be studied more quantitatively by the so called Lyapunov exponent. Lyapunov exponent 
describes sensitivity of the phase space trajectories to the initial conditions. It measures the growth rate between two initially nearby trajectories. \\
\begin{figure}[H]
\begin{center}
\includegraphics[width=.8\textwidth]{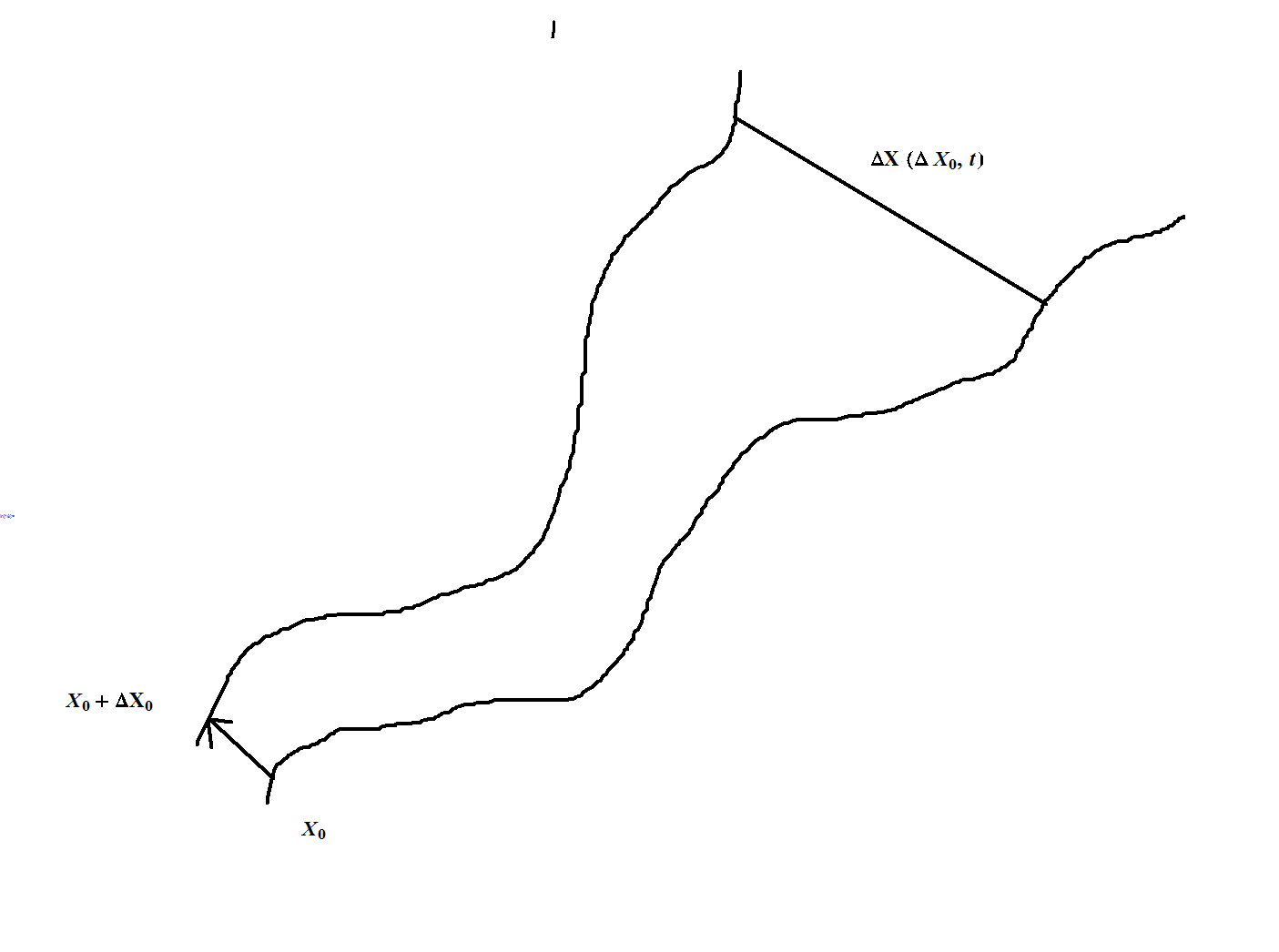}
    \caption{}
    \label{fig:f1}
\end{center}
\end{figure}
Let us consider two initially nearby orbits, one passes through the point $X_0$ and other $X_0+ \Delta X_0$. These orbits can be thought of as parametric functions of time. If $\Delta X(X_0,\tau) $ is the separation between these two orbits at a later time $tau$, then the Lyapunov 
exponent is defined as 
\begin{align}
\lambda = \frac{1}{\tau}ln \frac{\| \Delta X(X_0,\tau)\|}{\| \Delta X_0\|}
\end{align}
It will be useful to take the largest Lyapunov exponent which can be measured when the interval is very large.
\begin{align}
\Lambda = \lim_{\tau->\infty}\frac{1}{\tau}ln \frac{\| \Delta X(X_0,\tau)\|}{\| \Delta X_0\|} = \lim_{\tau->\infty} \frac{1}{\tau}\sum \lambda_i \tau_i
\end{align}
The largest Lyapunov exponent generally converges to a positive value for a physical system which exhibits chaotic behaviour. 
If $\Lambda$ is zero then it indicates the system is conservative. For a non conservative system or dissipative system the 
$\Lambda$ converges to a negative value \cite{Hiborn},\cite{sprott}.
\begin{figure}[t!]
    \centering
    \begin{subfigure}[t]{0.5\textwidth}
        \centering
        \includegraphics[width=1\textwidth]{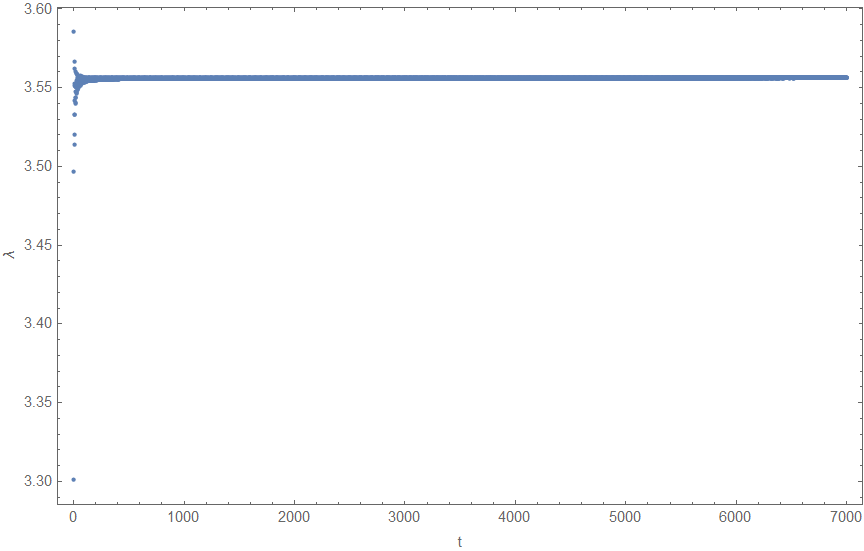}
        \caption{Extended string}
    \end{subfigure}%
    ~ 
    \begin{subfigure}[t]{0.5\textwidth}
        \centering
        \includegraphics[width=1\textwidth]{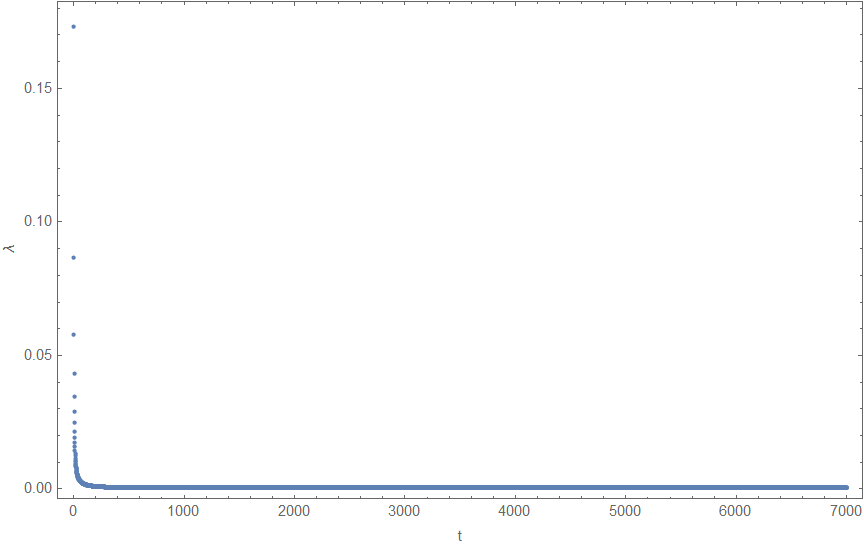}
        \caption{Point like string}
    \end{subfigure}
    \caption{Largest Lyapunav Exponents for (a) Extended string with initial condition $\theta_1(0)=0.1,p_{\theta_1}(0)=0,\theta_2(0)=0.1,p_{\theta_2}(0)=0.203444$ (b) point like string with initial condition $\theta_1(0)=0.1,p_{\theta_1}(0)=0,\theta_2(0)=0.1,p_{\theta_2}(0)=0.2041$}
    \end{figure}
For extended string the largest Lyapunav exponent Fig 6(a) converges to a positive value close to 3.55 which verifies extended string motion in $\gamma$ deformed $AdS_5 \times T^{1,1}$ is chaotic in nature. But for point like string Fig 6(b) its value converges to zero which 
indicates non-chaotic nature of the system. We arrived at this conclusion by studying the problem mostly numerically. 
    \newpage
    \section{Conclusion}
In this paper we have explicitly showed through the appearance of chaos that the string motion in the $\hat{\gamma}$ 
deformed $AdS_5 \times T^{1,1}$ geometry is not
integrable. We arrived at this conclusion by studying the problem mostly numerically. We study numerically the motion of the system and found 
it to be chaotic. Non-integrability does not, necessarily, imply the chaotic motion. However, the appearance of chaos is evidence of the breakdown
of integrability. In our study the chaotic motion of the strings is first seen in the Poincar\'{e} sections and also in the phase space trajectories. We have
taken the example of a particular type of circular string and showed that numerically that the motion is chaotic. We have shown further that as soon
as the strings are replaced with point particles the integrability restores back. Hence one can conclude that while the point like solutions are 
integrable, the extended string equations of motion are not. We have further support of this result by looking at the Lyapunov exponent and proved 
that the string equations of motion are non integrable while the point like equations are integrable. There are various things that one could look at. 
First whether there is any link between the marginal deformations and the chaos. Whether all marginal deformations lead to chaotic behaviour ?
Further one may ask what happens to these classical chaos at the quantum level. Of course the string trajectories will lead to excitations of heavy string states. Hence in the field theory side they would correspond to operators withe very large quantum numbers. Finally, the method of 
nonintegrability and chaos in classical string trajectories might help us better in understanding the  gauge/gravity duality better in terms of looking 
at how the nonintegrability really affects the field theory operators. \vskip .2 in\noindent {\bf Acknowledgement:} KLP would like to thank  
DESY theory group for hospitality under SFB fellowship where a part of this work is done.    

\end{document}